\documentclass[conference]{IEEEtran}
\IEEEoverridecommandlockouts
\usepackage{cite}
\usepackage{amsmath,amssymb,amsfonts}
\usepackage{algorithmic}
\usepackage{graphicx}
\usepackage{textcomp}
\usepackage{xcolor}
\usepackage{amsmath,graphicx, tabularx, siunitx}
\usepackage{tabularx, amssymb, pifont, arydshln}
\usepackage{cite}
\usepackage{hyperref}
\usepackage[absolute,overlay]{textpos}

\def\BibTeX{{\rm B\kern-.05em{\sc i\kern-.025em b}\kern-.08em
    T\kern-.1667em\lower.7ex\hbox{E}\kern-.125emX}}
\begin{document}

\begin{textblock*}{\textwidth}(1.8cm, 10.3in) 
    \noindent\footnotesize © 2025 IEEE. Personal use of this material is permitted. Permission from IEEE must be obtained for all other uses, in any current or future media, including reprinting/republishing this material for advertising or promotional purposes, creating new collective works, for resale or redistribution to servers or lists, or reuse of any copyrighted component of this work in other works.
\end{textblock*}

\title{Contrastive Knowledge Distillation for Embedding Refinement in Personalized Speech Enhancement\\}

\author{
    \IEEEauthorblockN{
    Thomas Serre\IEEEauthorrefmark{2}\IEEEauthorrefmark{3}\thanks{This work was performed using HPC resources from GENCI–IDRIS (Grant 2023-AD011014877).}, 
    Mathieu Fontaine\IEEEauthorrefmark{3}, 
    Éric Benhaim\IEEEauthorrefmark{2},
    Slim Essid\IEEEauthorrefmark{3}
}

\IEEEauthorblockA{
    \IEEEauthorrefmark{2}\textit{Signal Processing and Machine Learning departement}, \textit{Orosound}, Paris, France\ \\
      \IEEEauthorrefmark{3}\textit{LTCI}, \textit{T\'el\'ecom Paris}, \textit{Institut polytechnique de Paris}, Palaiseau, France\
    }
}


\newcommand{\se}[1]{{\textcolor{black}{#1}}}

\newcommand{\mf}[1]{{\textcolor{blue}{#1}}}
\newcommand{\mfcmt}[1]{{\textcolor{blue}{#1}}}

\newcommand{\secor}[2]{{\textcolor{black}{#2}}}
\newcommand{\secmt}[1]{}

\newcommand{\ts}[1]{{\textcolor{black}{#1}}}
\newcommand{\tscor}[2]{{\textcolor{black}{#2}}}
\newcommand{\tscmt}[1]{}

\maketitle
 
\begin{abstract}
Personalized speech enhancement (PSE) has \tscmt{recently} shown convincing results when it comes to extracting a known target voice among interfering ones.
The corresponding systems usually \se{incorporate a representation of the target voice within the enhancement system, which is} extracted from an enrollment clip of the target voice with upstream models. 
Those models are generally heavy as the speaker embedding's quality directly affects PSE performances. 
Yet, embeddings generated beforehand cannot account for the variations of the target voice during inference time. 
In this paper, we propose to perform on-the-fly refinement of the speaker embedding using a tiny speaker encoder. 
We first introduce a \se{novel} contrastive knowledge distillation methodology in order to train a 150k-parameter encoder from complex embeddings.
We then use this encoder \secor{for comparing the reference embedding with the projected input mixture}{within the enhancement system during inference and} show that the proposed method greatly improves PSE performances while maintaining a low computational load. 
\end{abstract}
\begin{IEEEkeywords}
Target speaker extraction, speech enhancement, speaker embedding, knowledge distillation
\end{IEEEkeywords}

\section{Introduction}

With the increase of telecommuting, improving remote communications and tackling privacy issues by preventing the leak of confidential background conversations has become a rising topic.
Though deep neural networks have led to superior speech enhancement (SE) results, such models cannot extract a target voice from overlapping interfering voices without supplementary information on this voice. 
Using additional cues on a target speaker, personalized speech enhancement (PSE) aims to extract the target speaker's voice among other interfering voices and/or environmental noises \cite{wang2018voicefilter, xu2020spex, vzmolikova2019speakerbeam, giri2021persopercepnet}. 
PSE systems usually comprise an upstream speaker encoder and a downstream SE network. 
The former generates a speaker embedding of the target speaker from an input enrollment clip, which is then fed to the downstream model to isolate the target voice from the noisy mixture.  
Both models are either trained jointly \cite{xu2020spex, ge2020spex+, ge2021spex++, ji2020speaker_joint},  or independently \cite{wang2018voicefilter, giri2021persopercepnet, newmodels_pdccrn}. 

PSE performances are thus highly correlated with the target speaker's embedding quality. 
Many speaker embedding \se{systems} have been proposed, such as i-vectors \cite{dehak2010ivector}, x-vectors \cite{snyder2018xvector}, d-vectors \cite{wang2018voicefilter}, or ECAPA-TDNN embeddings \cite{ecapa}. 
Since identifying the target voice in a noisy content is closely related to the Speaker Verification (SV) task, most encoders are trained for this task beforehand.
A contrastive approach is usually used in order to group together embeddings of the same speaker and push apart the ones of different speakers.  
In the quest for lighter speaker encoders, Knowledge Distillation (KD) stands as a relevant approach.
This technique aims to train a smaller \textit{student} network to mimic the \secor{results}{outputs} of a bigger \textit{teacher} model. 
Such \se{an} approach was proposed in \cite{ecapa_lite}, leading to a light version of the \tscmt{well-known} ECAPA-TDNN system \cite{ecapa}. 

When it comes to the downstream PSE model, several options have been proposed, either adapting existing SE systems \cite{newmodels_pdccrn, giri2021persopercepnet}, or directly proposing PSE systems \cite{wang2018voicefilter, thakker2022fast}.
Dual stage frameworks have recently shown great performance, both in SE \cite{2stage} and PSE \cite{ju2022tea}, paving the way for new architectures. 
Despite those new powerful architectures, the speaker's voice at inference time may differ from that of the enrollment. 
Thus, having an embedding that perfectly matches the target voice at inference time remains a crucial issue.
Recent works have focused on adapting the speaker embedding at inference time. 
For example, \cite{adaptive_speakerbeam} proposed to add spatial features that allow for on-the-fly adaptation, but require multi-channel signals. 
Other systems proposed iterative approaches \cite{it_refined, multistage} that consist in reusing the enhanced mixture in order to refine the embedding thus increasing the computational cost. 

In this work, we propose to train a lightweight speaker encoder using a contrastive knowledge distillation strategy 
and use it to adapt the target speaker's \secor{information}{representation} to the input mixture \secor{in real-time}{on-the-fly}. 
We first detail the proposed training approach \secor{which the}{whose} goal is to match low footprint speaker embeddings with high complexity speaker embeddings. 
We show that this contrastive training strategy allows for the use of a 100 times smaller encoder while maintaining sufficient speaker information.  
Then, based on this tiny encoder, we propose a simple scheme for on-the-fly speaker adaptation of the embedding. 
This is achieved by computing the similarity between the reference embedding and light embedding of the input noisy frames.
Our results show that using this similarity as an additional cue greatly improves the PSE performances.




\section{Proposed Method}

In this section, we detail the proposed tiny speaker encoder and its contrastive KD training. Then, we present the PSE framework as well as our novel embedding refinement strategy. 

\begin{figure*}[t]
    \centering
    \includegraphics[width=0.89\linewidth]{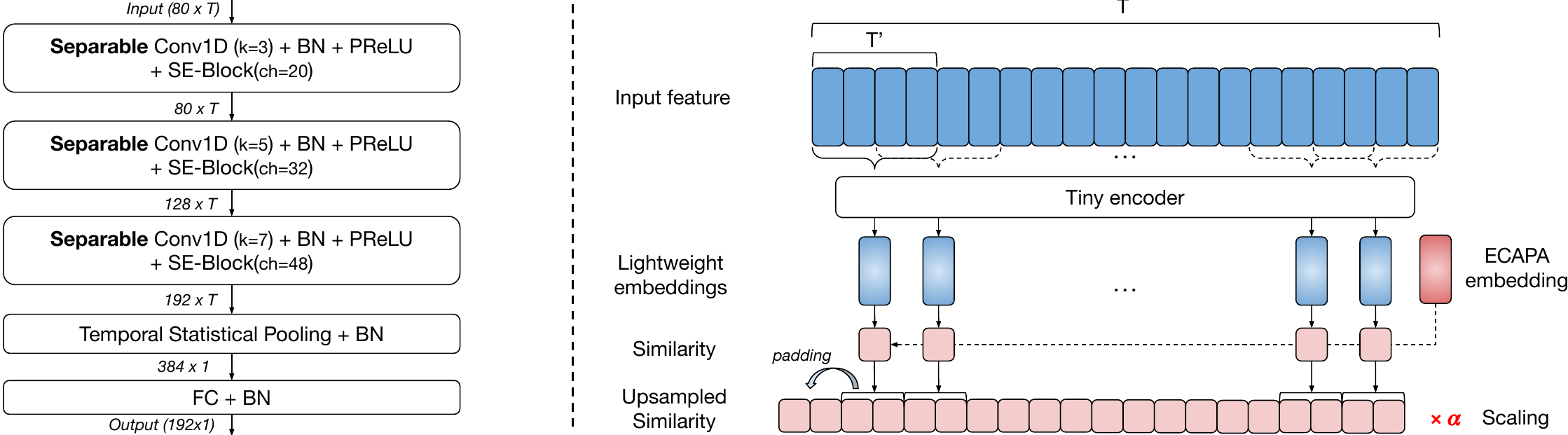}
    \caption{(Left) Tiny encoder's architecture (Right) Methodology for on-the-fly similarity computation. \secmt{wrong order corrected. Was L introduced in the text?} }
    \label{fig:archi_sim}
\end{figure*}

\subsection{Speaker embeddings}


\noindent
\textbf{ECAPA-TDNN.} Inspired by the x-vectors \cite{snyder2018xvector}, ECAPA-TDNN \cite{ecapa} uses Time-Delay Neural Networks combined with Squeeze-Excitation modules \cite{ecapa} to extract the \secor{acoustic information}{speaker information} from input Mel-spectrogram features. An attentive pooling layer is then used to embed the \secor{temporal information}{speaker representation} into a fixed-size vector. The Additive Margin softmax loss \cite{ammloss} is used for training, which ensures optimal embedded representations. Having achieved state of the art performance in SV, this system is now widely used in PSE as it yields speaker-discriminating embeddings. In this work, we use ECAPA-TDNN as a baseline and a teacher for the KD approach. 

\noindent
\textbf{TinyECAPA.} In this paper, we propose a tiny speaker encoder \secor{inspired by}{whose design is inspired by} previously proposed heavy systems \cite{ecapa}. 
It consists of 3 convolution blocks, a temporal pooling layer, and a projection linear layer \se{(as depicted in Fig. \ref{fig:archi_sim})}. 
The convolution block corresponds to a Separable Conv1D layer, a BatchNorm, a PReLU activation and a Squeeze-Excitation layer \cite{ecapa}.  
The temporal pooling is done by concatenating the mean and the standard deviation of the convolutions outputs over $T'$ frames.\secmt{of xxx over xxx temporal windows...} 
Finally, a linear layer coupled with a BatchNorm is used as a projector.
The input feature corresponds to the MFCC (excluding $C_0$) 
concatenated with delta and delta$^{2}$ along the frequency axis. Using this \se{extended} feature \se{vector} showed better performance for a smaller model size compared to MFCC only. 
Compared to ECAPA-TDNNLite, we do not use Res2Blocks nor attentive statistic pooling as we aim for minimal computational complexity. The obtained model has \se{only} 150k parameters. 

\subsection{Contrastive training strategy}

Our objective is to compare the reference embedding (computed offline) with the projected input mixture (computed at inference time). 
Thus, we propose a contrastive KD strategy that aims at matching the light embedding with the reference heavy embedding. We select a contrastive approach to benefit the most out of the speaker-discriminant asset of ECAPA-TDNN embeddings. \secmt{Before going futher, start with a sentence on motivation, why choose contrastive...}
Here, the projected input mixture is a sequence that we denote \se{by} $M = f_{\mathrm{enc}}(Y)$, where $f_{\mathrm{enc}}$ is the tiny encoder, $Y \in R^{N \times (3C-1) \times T}$ the input mixture feature, $M \in R^{N \times d_{E}\times (2\frac{T}{T'}-1)}$ the projected input sequence, $N$ the batch size, $C$ the number of MFCC coefficients, $d_{E}$\secmt{choose between $d_e$ and $d_F$} the embedding dimension, and $T'$ the temporal window used to generate one embedding, resulting in $2\frac{T}{T'}-1$ embeddings for the whole sequence as we use an overlap of $\frac{T'}{2}$.
Inspired by Contrastive Language-Audio Pretraining (CLAP)\secmt{define the acronym} \cite{clap}, we then construct a constrastive loss based on the similarity matrix.
To do so, we compute the cosine similarity for each frame $t'$ and then perform temporal pooling as follows: 
\begin{equation}
    S_{i, j} = {\frac{1}{T'}}\sum_{t'=0}^{T'}{E_{i}\cdot M_{j,t'}^{\top}}
\end{equation}
where $E_{i} \in R^{N \times d_{E}}$ corresponds to the reference ECAPA-TDNN embedding, $i$ and $j$ to the embedding index in the batch.\secmt{define K and Mj,t} The similarity matrix is thus $S = (S_{i,j})_{1 \leq i,j \leq N}$.
We then multiply $S$ by a temperature value \se{$\tau$} learnt during training and we use the following cost function $\ell = {\frac{1}{N}}\sum_{i=0}^{N}{\mathrm{log}(\mathrm{diag}(\mathrm{softmax}(\tau \cdot S)))}$. 
We do this for both axes resulting in two losses $\ell_{\mathrm{ECAPA}}$ and $\ell_{\mathrm{MFCC}}$ which we average to obtain the final loss $\mathcal{L}$.

\subsection{PSE framework: pDeepFilterNet2}
\label{section:pse_dfnet}

For the PSE downstream model, we select pDeepFilterNet2 \cite{pdfnet}, a lightweight dual-stage PSE system based on DeepFilterNet2 \cite{schroter2022deepfilternet2}, which is personalized by concatenating\secmt{say how} the speaker embedding with the encoder output features (see original paper \cite{pdfnet} for the architecture's details). 
We select the unified encoder version as we aim for optimal complexity, and we remove the linear layer used for local SNR estimation allowing for the model size reduction.
As in \cite{pdfnet}, the training loss is a combination of the spectral loss $\mathcal{L}_{\mathrm{spec}}$, the oversuppression loss $\mathcal{L}_{\mathrm{OS}}$, and the multi-resolution loss $\mathcal{L}_{\mathrm{MR}}$ using the 4 following window lengths: \{5, 10, 20, 40\} in ms. 

\subsection{On-the-fly embedding refinement}
\label{onthefly}

We take advantage of the low computational cost of the proposed speaker encoder to perform on-the-fly refinement \se{of the target voice characterisation} (see Fig. \ref{fig:archi_sim}). 
We do so by comparing a reference ECAPA-TDNN embedding with the embedded input mixture. 
To keep the temporal information of the input mixture we generate an embedding every chunk $T'$ of input frames with a 50\% overlap.
The output similarity is thus of size $2\frac{T}{T'}-1$, which is then upsampled and padded to match the initial number of frames.
The obtained similarity, which acts as a speaker activation detection, is  concatenated to the $T$-times repeated reference speaker embedding $E$. The refined embedding is finally fed to the PSE system following the integration methodology presented in  Section \ref{section:pse_dfnet}.

Though the cosine similarity lies in $[0, 1]$, when comparing lightweight embeddings and heavy embeddings, the resulting similarity rarely reaches 1 (insights given in Section~\ref{section:pse_results})\secmt{suggest an explanation why}.
To compensate for this, we introduce a scaling factor $\alpha>0$ that we multiply with the similarity before clipping \secor{the similarity between}{it in the range} $[0, 1]$.
This was found to be more efficient than more sophisticated approaches like $f(x)=\mathrm{sigmoid}(\alpha x + \beta)$.

\section{Experiments}
In this section, we present the experimental setups for both the training of the lightweight encoder and the PSE system.
\subsection{Speaker embedding training}
\noindent
\textbf{ECAPA-TDNN.} We use a pretrained ECAPA-TDNN\footnote{\url{https://github.com/TaoRuijie/ECAPA-TDNN/tree/main}}, trained on VoxCeleb2 and achieving an EER of 0.96\% on the VoxCeleb1-O test set with an embedding dimension of 192. For PSE training, we randomly select 15~s of the enrollment clip of each speaker to generate the embedding, which acts as an augmentation and allows for better generalization.

\noindent
\textbf{TinyECAPA.} Similarly, we train the tiny encoder with VoxCeleb2. 
We keep 5\% of the development speakers for validation. 
We construct pairs of enrollment clips of the same speaker where the first clip is fed to frozen ECAPA-TDNN and the other is projected with the tiny encoder with a chunk length $T'$ set to 1~s. 
We use 3-s excerpts and perform the same data augmentation as \cite{snyder2018xvector} which consists in mixing excerpts from the MUSAN dataset (babble noise, music) \cite{musan} and applying reverb from the RIR dataset in \cite{rir_dataset}. 
The number of coefficients for the MFCC is set to 27. Removing $C_0$ and adding delta and delta² leads to an 80-dimensional input feature \se{vector}.
The channels of the convolution blocks are set to [80, 128, 192], the kernels of the separable convolution to [3, 5, 7] and the channels of the SE-Blocks to [20, 32, 48]. Finally, the FC layer reduces the pooled embedding from 384 to 192.  

\begin{figure}[t]
    \centering
    \includegraphics[width=\linewidth]{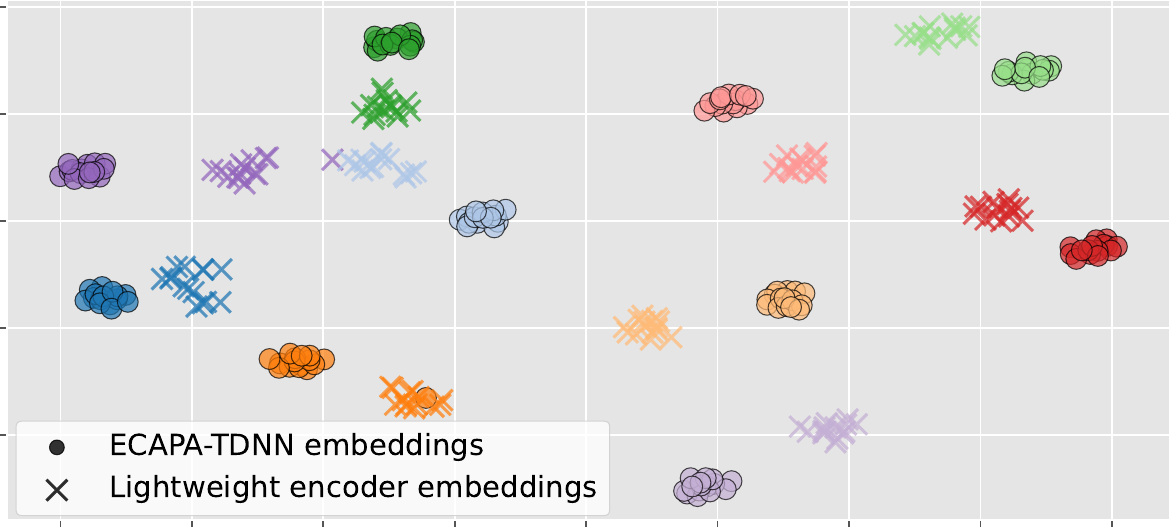}
    \caption{tSNE projection of 10 speakers' embeddings unseen during training, each represented by a color.}
    \label{fig:tsne}
\end{figure}

\subsection{PSE training}

\noindent
\textbf{Dataset.} For the PSE training set, we use the DNS5 dataset \cite{dubey2023icassp}. We use 10 s of the source files for target speech and interfering speech, and the whole files for noise and RIRs. 
We also use a small corpus
of Mozilla common voice \cite{ardila2019common} to improve interference diversity. 
The excerpts are generated on-the-fly and 15\% of the source files are kept to generate a fixed 50-h validation set with the same settings as  the training set.
The generated samples are either \textit{target+noise} (20\%), \textit{target+interference} (30\%) or \textit{target+interference+noise} (50\%).
The SNR and the SIR values are drawn using a truncated Gaussian distribution as in \cite{pdfnet}.

\begin{table*}[t]
    \caption{PDNSMOS results on the DNS5 blind test set$^*$. The higher the better. Underlined: best of all results \\
    and bold: best of the proposed systems. Statistical relevance ensured by Wilcoxon tests ($p<0.05$).
    }
    \centering
    {\renewcommand{\arraystretch}{1.1}%
    \begin{tabularx}{0.90\linewidth}{l*{2}{c}|*{3}{c}|*{3}{c}|*{2}{c}}
    \hline
    Model & \multicolumn{2}{c}{Speaker Embeddings} & \multicolumn{3}{c}{Track 1: Headset} & \multicolumn{3}{c}{Track 2: Speakerphone} & \multicolumn{2}{c} {Complexity}   
    
    \\\hline
          & Reference & On-the-fly & SIG  & BAK   & OVRL        & SIG    & BAK   & OVRL & Params (M) & MACs (G)      \\\hline
    Noisy & - & - & \underline{\num{4.152}} & \num{2.369} & \num{2.709} & \underline{\num{4.0460}} & \num{2.159} & \num{2.497} &-&-\\ 
    

    E3Net & ECAPA & - & \num{3.824} & \num{3.474} & \num{3.096} & \num{3.716} & \num{3.167} & \underline{\num{2.872}} & 6.62 &  \num{11.07711}$^\mathrm{**}$\\ 

    pDCCRN & ECAPA & - & \num{3.763} & \num{3.290} & \num{2.985} & \num{3.650} & \num{3.090} & \num{2.814} & 4.50 & \num{5.589967} \\ \hline \hline

    pDeepFilterNet2 & ECAPA & - &  \num{3.77276}  & \num{3.47003} & \num{3.07800} & \num{3.580214}& \num{3.17067} & \num{2.77719} & 2.23 & 0.33\\ 
    
    \hspace{2mm}- \textit{oracle similarity}& ECAPA & ECAPA & \textbf{\num{3.85323}}&\num{3.32697}&\num{3.07958}&\textbf{\num{3.64455}}&\num{3.04442}& \num{2.76860}& \num{17.6745} & \num{1.7527}\\

    \hspace{6mm} \textit{w/o scaling} & ECAPA & ECAPA  & \num{3.81741}  & \num{3.43359} & \num{3.10597}  & \num{3.5933}& \num{3.12141} & \num{2.77699} & \num{17.6745} & \num{1.7527} \\
    
    \hspace{2mm}- \textit{light similarity} & ECAPA & TinyECAPA &  \num{3.80966}  & \num{3.49174} & \underline{\textbf{\num{3.12600}}}  & \num{3.63237}& \num{3.18519} & \textbf{\num{2.82878}} & 2.38 & \num{0.374} \\
    
    \hspace{6mm} \textit{w/o scaling} & ECAPA & TinyECAPA &  \num{3.66912}  & \underline{\textbf{\num{3.60272}}} & \num{3.06257}  & \num{3.52582}& \underline{\textbf{\num{3.30025}}} & \num{2.814337} & 2.38 & \num{0.374} \\

    \hline
    \multicolumn{11}{l}{ \textit{{\hspace{-2pt}\footnotesize($^\mathrm{*}$) Samples available at \url{https://pdeepfilternet2.github.io/} in Part II. ($^\mathrm{**}$) The learnable decoder takes up to 10.5G MACs}}} \\


    \end{tabularx}} \quad
    
    \label{tab:pse_results}
\end{table*}

\noindent
\textbf{Training setup.} Compared to \cite{pdfnet}, we use 16 kHz as sample rate which allows us \se{to target our application scenario and} to reduce the number of ERB to 24. All other parameters remain the same.
We perform early stopping on the validation loss with a patience of 20 epochs and the loss factors are found during training: $\lambda_{\mathrm{spec}}=1e3$,  $\lambda_{\mathrm{MR}}=5e2$ and $\lambda_{\mathrm{OS}}=5e2$ respectively for $\mathcal{L}_{\mathrm{spec}}$, $\mathcal{L}_{\mathrm{MR}}$ and $\mathcal{L}_{\mathrm{OS}}$.\secmt{always say what procedure was used to find hyp. param. values}

\section{Results}
\label{results}
In this last section, we present our results. We first evaluate the proposed tiny encoder on a SV task, before showing the added value of the similarity refinement for the PSE task.
\subsection{Speaker encoders}

We first confirm the performance of the tiny speaker encoder on a SV task.
We evaluate on the VoxCeleb1-O test set using the Equal Error Rate (EER) metric and the minimum normalized detection cost MinDCF \cite{voxsrc} with $P_{\mathrm{target}}=10^{-2}$ and $C_{\mathrm{FC}}=C_{\mathrm{Miss}}=1$. 
We consider the ECAPA-TDNNLite asymmetric enroll-verify system for comparison \cite{ecapa_lite}. Precisely, we compare the last row of Table 2 in \cite{ecapa_lite}  which corresponds to the use of ECAPA-TDNN for enrollment and ECAPA-TDNNLite for verification. Similarly, we use ECAPA-TDNN for enrollment and TinyECAPA for verification. 
ECAPA-TDNNLite achieves an EER of 2.31\% and a MinDCF of 0.251 for a model size of 318k parameters. With TinyECAPA, we obtain an EER of 4.26\% and a MinDCF of 0.337 for a model size of 150k parameters. 
We have thus managed to greatly reduce the size of our encoder while keeping acceptable performance.
Moreover, our training strategy only requires a pretrained ECAPA-TDNN model thus saving training resources.
We also project TinyECAPA embeddings along with ECAPA-TDNN embeddings using the tSNE algorithm for 10 speakers in Fig. \ref{fig:tsne} which showcases the performance of the KD approach.



\subsection{PSE evaluation setup}
\label{section:pse_results}

\noindent
\textbf{Evaluation set.} We assess our frameworks using the DNS5 Blind test set, which features real data, to test out-of-domain generalization.  
This set is separated into "Headset" and "Speakerphone" tracks.
The latter is harder as the target voice is further from the microphone than with Headset samples. 

\noindent
\textbf{Metrics.} We select the PDNSMOS \cite{reddy2021dnsmos} as evaluation metrics as they do not require a clean reference.
The target extraction quality \se{is assessed} with the SIG metric and the background removal with the BAK metric. We also compare the overall performance with the OVRL metric. 

\noindent
\textbf{Baselines.} We compare our system to pDCCRN \cite{newmodels_pdccrn} and E3Net \cite{thakker2022fast}. 
We implemented pDCCRN based on publicly available implementation of DCCRN \cite{dccrn_git}, following the instructions of \cite{newmodels_pdccrn}.
We trained the system with the same architecture and power-law compressed phase-aware asymmetric loss as in \cite{newmodels_pdccrn}.
For E3Net, we implemented our version based on the details provided in \cite{thakker2022fast}, keeping the same architecture's settings and selecting the \textit{baseline} model ($N\!=\!4$). 
As in \cite{quant_emb_study}, we used the SI-SNR loss \cite{sisdrloss} as training objective.
Both models were trained for 600k iterations with the Adam optimizer and a cosine learning rate scheduler.
We respectively set the peak learning rate to $10^{-3}$ and $10^{-4}$ for pDCCRN and E3Net. 

\noindent
\textbf{Experiments.} In Table \ref{tab:pse_results} we evaluate pDeepFilterNet2 as baseline and four similarity-based systems using two distinct similarity refinement \se{strategies}: the \textit{oracle} similarity and the \textit{light} similarity.
The first strategy encodes the input mixture with ECAPA-TDNN while the second strategy encodes it with the tiny encoder. 
Both methods encode the reference embedding with ECAPA-TDNN.
For both strategies we propose an unscaled version and a scaled version. In the latter, the scaling factor $\alpha$ is set as the value that brings the better trade off between the SIG and the BAK metrics. 
We set it to $2$ for the \textit{oracle} similarity and to $6$ for the \textit{light} similarity.


\setlength{\abovecaptionskip}{1pt}

\begin{figure}
    \centering
    \includegraphics[width=0.95\linewidth]{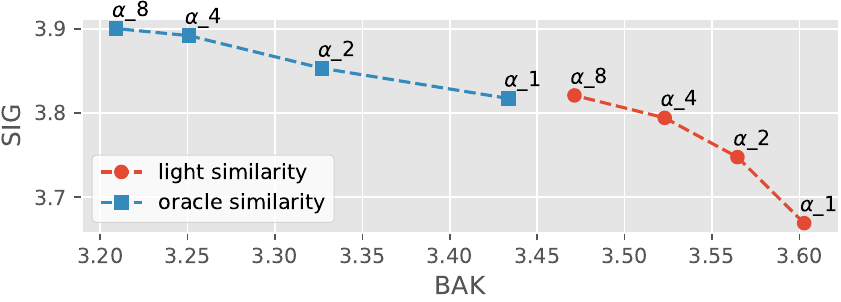}
    \caption{SIG vs BAK on Track1 of DNS5 Blind test set for different values of the scaling factor $\alpha$}
    \label{fig:enter-label}
\end{figure}

\subsection{Discussions}

\noindent
\textbf{Similarity refinement.}
We first see that all similarity-based versions improve at least one metric compared the pDeepFilterNet2 baseline.
Yet, the \textit{oracle} similarity improves the SIG metric at the cost of the BAK metric. This can be explained by the temporal support $T'$ selection. Indeed, with the attentive pooling layer, ECAPA-TDNN is more prone to False Positives with a long  $T'$ compared to the tiny encoder which uses standard temporal pooling. 
On the contrary, the \textit{light} similarity is better at improving the BAK metric. 
In the end, for our chosen $T'$, the \textit{light} similarity achieves better performances than the \textit{oracle} similarity with the appropriate scaling.

\noindent
\textbf{Scaling factor.} As the ablation study shows, removing the scaling greatly impacts the similarity-based systems. Acting as a speaker activation, having a small range of variation makes it difficult to identify when the target speaker is active. This is especially the case for the \textit{light} similarity as it does not span the whole $[0, 1]$ range\secmt{comment ca? tu veux dire doesn't span the whole range?}. This can be explained with Fig. \ref{fig:tsne}: light embeddings and heavy embeddings are close to each other but never overlapped. 
The scaling factor is also directly linked to the trade off between the signal quality and background removal as one can see on Fig.~\ref{fig:enter-label}. 
Indeed, increasing $\alpha$ increases the True Positive Rate (TPR) while diminishing the False Negative Rate (FNR), and decreasing the $\alpha$ diminishes the TPR while increasing the FNR. 

\noindent
\textbf{Baselines comparison.}
Finally, we compare our system to previously proposed PSE models. 
Our baseline model achieves similar performance as pDCCRN and lower performance compared to E3Net. Yet, when improved with similarity refinement, our system reaches performance comparable to E3Net, especially on Track 1 with the \textit{light} similarity. Moreover, our system is much smaller and its complexity in MACs is greatly smaller than the one of both pDCCRN and E3Net. 
In the end, the similarity refinement can be seen as a deported module that alleviates the PSE model in the speaker activation detection task. This allows for a greater trade off between extraction performances and model complexity, making our system competitive with an almost 3 times bigger model.

\section{Conclusions}

In this work, we proposed a novel approach for on-the-fly speaker embedding refinement in PSE. 
We first trained a tiny speaker encoder using a contrastive KD methodology that allowed us to benefit from ECAPA-TDNN deep representations. 
Achieving unprecedented reduction in model size, we took advantage of this low complexity to compute the cosine similarity between the reference embedding and the input mixture's embedding at inference time. 
We showed the added value of the proposed refinement compared to the baseline and previously proposed systems. 
Future work includes an improved scaling methodology as well as the use of the contrastive KD approach on other SV systems.

\newpage
\bibliographystyle{IEEEtran}
\bibliography{mybib}

\end{document}